# Conditions and possible mechanism of condensation of *e-h* pairs in bulk GaAs at room temperature

Peter P. Vasil'ev

*PN Lebedev Physical Institute, 53 Leninsky prospect, Moscow 119991, Russia;*

peter@mail1.lebedev.ru

A mechanism of the condensation of *e-h* pairs in bulk GaAs at room temperature, which has been observed earlier, is proposed. The point is that the photon assisted pairing happens in a system of electrons and holes that occupy energy levels at the very bottoms of the bands. Due to a very high *e-h* density, the destruction of the pairs and loss of coherency does not occur because almost all energy levels inside a 30-60 meV band from the bottom of the conduction band prove to be occupied. As a result, a coherent ensemble of composite bosons (paired electrons and holes) with the minimum possible energy appears. The lifetime of this strongly non-equilibrium coherent *e-h* BCS-like state is as short as a few hundred of femtoseconds.

**1. Introduction**

Bose condensation of excitons has been one of the most fascinating and rapidly growing fields of modern condensed matter physics during the last years. It is well known that electrons and holes in semiconductors, being fermions, can form bound (paired) states, which exhibit properties of Bose statistics [1]. In case of low temperatures and densities, such bound states are referred to as excitons. Bose condensation of excitons has been previously observed at cryogenic temperatures in some semiconductors, for example, in $Cu_2O$ and quantum-well GaAs/AlGaAs heterostructures [2-5].

In the high-density limit, collectively paired electrons and holes behave like Cooper pairs in a superconductor, and the BCS-like energy gap at the Fermi level is the order parameter of the macroscopic quantum state [1,6]. As noted before [7], the order parameter of the excitonic condensate is identical to its optical polarization and can be directly accessed by relatively simple optical measurements. Moreover, it has been theoretically predicted that the dephasing and relaxation kinetics of excitonic condensate depends on its density in some cases [7]. In particular, the polarization dephasing rate slows down with increasing density of the condensate. This implies that the destruction of coherency due to collisions does not happen despite of an enhanced collision rate with increasing density.

In our recent experiments [8-11], we have studied the regime of the cooperative recombination in a highly non-equilibrium large density ($> 3 \cdot 10^{18}$ cm$^{-3}$) system of electrons and holes in bulk GaAs at room temperature. The main information about the properties of the electron-hole system was obtained from the analysis of emission spectra and their approximation in a similar manner like elsewhere [2,3,5]. It has been demonstrated that the photon-mediated collective pairing of electrons and holes and their condensation resulted in the formation of a short-living coherent *e-h* BCS-like state. Its radiative recombination was observed in a form of very powerful femtosecond optical pulses. It has been experimentally demonstrated that almost all electrons were condensed at the bottom of the conduction band and were in a collective state [11]. An average lifetime of the *e-h* BCS-like state was measured to be around 300 fs.

The most unclear and unexpected facts were the condensation at room temperature, the stability of the condensate, and the conservation of coherency of the *e-h* ensemble for much longer times than the intraband relaxation time. Here, a possible explanation of these facts is proposed. Let us point out two fundamental conditions of the *e-h* condensation at room temperature from the very beginning. First, the presence of a resonant electromagnetic field with the quantum energy equal to *e-h* transitions between the bottoms of the conduction and valence bands is essential. Second condition is a very large *e-h* density that, due to the Pauli exclusion principle, results in the occupation of almost all energy levels in a wide enough band above the bottom of the conduction band. These two conditions must be fulfilled simultaneously and the absence of any of theses makes the condensation unfeasible.

In the present paper, an additional experimental evidence of the condensation is presented. It will be demonstrated that a very large peak power of the emission, which is generated from a very narrow spectral region at a femtosecond duration, can not be explained by a standard Fermi-Dirac distribution of electrons and holes in the bands and the square-root dependence of the density of states.

**2. Experiment**

The measurements were carried out at room temperature. We used similar structures as described elsewhere [8-11]. A characteristic feature of the structures is the presence of two sections along the axis of the device, one being an optical amplifier, while the other being an absorber. In addition, there exist a relatively poor quality resonator in the structure comprised by the cleaved facets of the chip. Large amplitude (500-800 mA) nanosecond current pulses were applied on the amplifier section, whereas the absorber was connected to a d.c. reverse bias source. Figure 1 (a) shows a typical current waveform. The repetition rate of the pulses was 39.74 MHz, the pulsewidth was less than 10 ns.

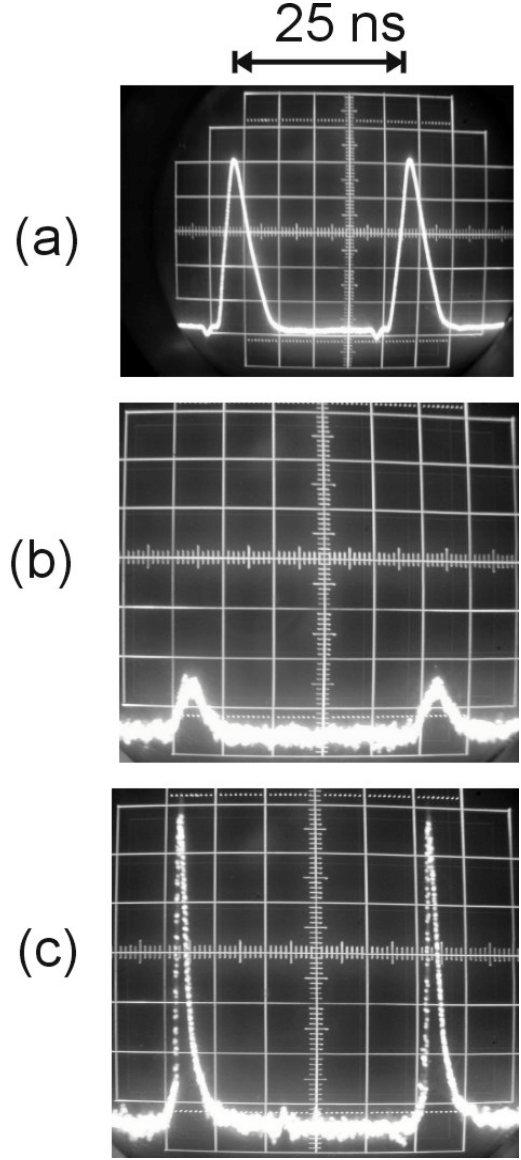

Fig 1. Waveforms of the current pulses (a), amplified spontaneous emission (b), and cooperative emission (c).

The net gain or net loss was readily achieved in the structure by changing the pumping of both the amplifier and absorber sections (see Section 4). For instance, lasing can be easily achieved in the structures at driving current pulses of around 100 mA and more and zero reverse bias on the absorber section. On the other hand, it is possible to shift the absorption edge towards the longer wavelength and suppress the optical gain at all frequencies by the application of a larger reverse bias on the absorber (the Franz-Keldysh effect). The lasing is suppressed in this case and the amplified spontaneous emission is generated. Here, the output emission looks like a train of long (nanosecond) pulses with a peak power of a few hundreds μWs. The spontaneous emission pulses are shown in Fig. 1 (b). They were detected by a *p-i-n* photodiode with a time resolution of 400 ps and a 20-GHz bandwidth sampling oscilloscope.

However, if one increases the pumping to the amplifier section at the same reverse bias, the optical gain at the very long-wavelength edge of the spectrum overcomes the absorption due to the band filling effect and the band gap shrinkage. As a result, photons, with quantum energy equal to the intraband transition between the bottom of the conduction band and the top of the valence band, can travel freely through the structure and establish coherency in the electron-hole system. As noted before [8,9], the cooperative emission can be generated. Fig. 1(c) illustrates ultrashort pulses of the cooperative emission, which were detected with the photodiode. Since the actual duration of the pulses lies in the femtosecond region, the detected pulsewidth observed on the oscilloscope screen corresponds to the temporal resolution of the photodiode.

A standard autocorrelation technique based on the second harmonics generation (SHG) has been utilised for more precise measurement of the pulsewidth [12]. Figure 2 shows a typical fringe-resolved autocorrelation function of the cooperative emission. The full width at half maximum (FWHM) of the SHG trace is about 350 fs. This corresponds to the actual pulsewidth of 230 fs assuming the pulseshape in the form of the asymmetric *sech* [12].

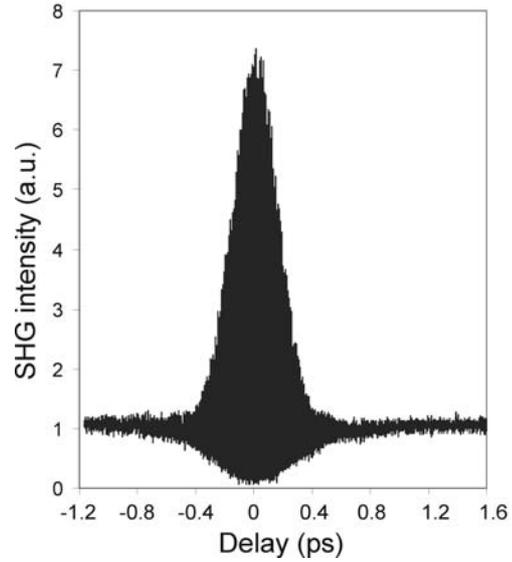

Fig. 2. SHG autocorrelation function of superradiant pulses. The pulsewidth is 230 fs.

Spectral characteristics of the emission were assessed using a monochromator with a resolution of 0.1 nm. The optical spectrum of the amplified spontaneous emission is very broad (see Fig. 3 a). Its width is about 52 meV at the driving current of 730 mA and the reverse bias of – 6.3 V.

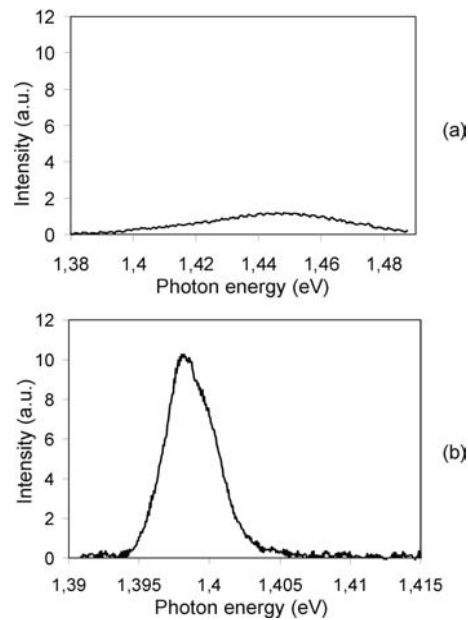

Fig 3. Optical spectra of amplified spontaneous emission (a) and cooperative emission pulses (b). Note different horizontal scales in (a) and (b).

The central wavelength is 858 nm, which corresponds to the photon energy of 1.448 eV. The cooperative emission spectrum at the same driving current and the reverse bias of – 6.1 V is shown in Fig. 3 (b). It is much narrower (the FWHM of 4 meV) and its peak wavelength is shifted strongly towards longer wavelengths. The photon energy of the spectral peak is 1.398 eV. The whole spectrum lies at the very edge of the renormalised band gap, which corresponds to the energy of $E_g$ = 1.394 eV at the estimated carrier density level of around $6\,10^{18}$ cm$^{-3}$ [10].

The peak power of the cooperative emission pulses exceeds that of both amplified spontaneous emission and lasing by a few orders of magnitude. Here, the pulse energy is of the same order as the energy of the spontaneous emission background between the pulses. It was found that the energy of the femtosecond cooperative pulses was slightly larger at smaller repetition rates (< 1MHz). Its typical value was around 30 pJ, whereas the maximum value exceeded 40 pJ. Estimated peak powers were in a range of 100-150 W at pulsewidths of 230-280 fs. These values are absolute records for pulses generated by this particular type of semiconductor structures.

It is very important for us that the very large optical power is generated during a very short time from a very narrow spectral width that locates at the edge of the band gap. Let us now demonstrate that this is possible only when a highly non-equilibrium condensed e-h state is formed near the band gap during the collective emission generation.

## 3. The maximum possible power and density of states

In this Section, we show that the extremely large peak power (energy) generated from a narrow spectral range at femtosecond pulsewidths cannot be explained by the standard Fermi-Dirac distribution of electrons and holes in the bands and the square root dependence of density of states (DOS). Suppose that the generation of the cooperative emission pulses, which are presented in Figs. 1-3, is determined by the radiative recombination of conventional electron-hole plasma. Let us estimate the maximum possible energy and peak power of the emission for the ideal case, when all *e-h* pairs recombine simultaneously and transform into photons without loss. Obviously, actual values of the pulse energy and power are smaller.

Let consider a given spectral range, say, 8 meV inside the conduction band near the band gap. One can see in Fig. 3 that this value is approximately equal to the experimental bandwidth where about 90% of the total energy of the pulses locate. Figure 4 shows the standard dependence of DOS on energy, which is proportional to $\sqrt{E - E_g}$. We will not take into account the long wavelength tail of DOS at $E < E_g$, because the portion of energy of the pulses in this range is negligible, as one can notice in Fig. 3 (b).

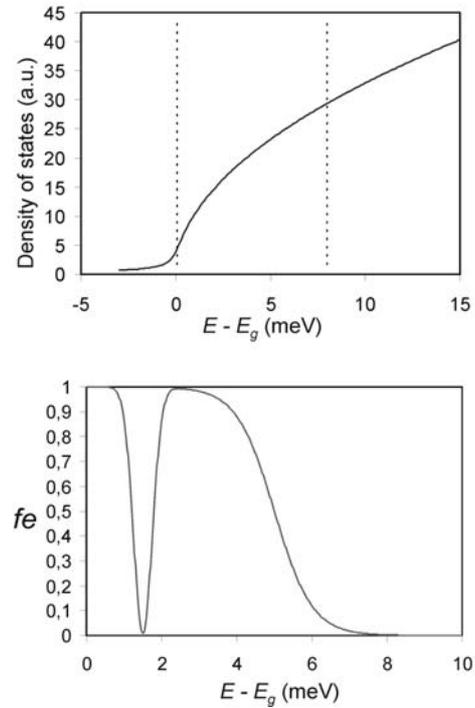

Figure 4. Density of states of electrons and the spectral hole in the energy distribution after their simultaneous recombination.

The electron concentration at the energy region shown in Fig. 4 is given by

$$n = \int_{E_g}^{E_g+E_1} N_c(E) f_e(E) dE \quad (1)$$

where $N_c(E) = (2m_e)^{3/2}(E-E_g)^{1/2}/(2\pi^2\hbar^3)$ is the density of states, $f_e(E) = \{\exp[(E-E_g-\mu)/kT]+1\}^{-1}$ is the Fermi-Dirac distribution, $\mu$ is the Fermi energy, $k$ is the Boltzman constant, $T$ is the temperature (= 300 K), $E_1$ = 8 meV. Because in our case of high pumping rates and large carrier densities the electrons are degenerated, we suppose for simplicity that $f_e = 1$, i.e. all the energy levels within the range of $E_g < E < E_1$ are occupied. Then, the electron concentration is $n = 4.3\times10^{17}\left(\frac{8}{25}\right)^{3/2} \approx 7.8\times10^{16}$ cm$^{-3}$.

The number of electrons, which take part in the radiative recombination, is obviously equal to $N = n \times V$, where $V$ is the volume of the pumping region. We have $V$ = 200 by 0.13 by 7 μm = 1.82·10$^{-10}$ cm$^3$. Consequently, N = 1.4·10$^7$ electrons. In the ideal case of the maximum possible power all the electrons recombine with holes and produce the same number of photons. Since the semiconductor structure has two optical outputs, about $N_{ph}$ = 7·10$^6$ photons are emitted from each facet of the chip. The output pulse energy is

$$E_p = \hbar\omega N_{ph}(1-R) \quad (2)$$

where $\hbar\omega$ is the quantum energy and $R$ is the reflectivity of the facet. This is the maximum possible energy that can be generated within the energy band of 8 meV just above the band gap.

With the experimental values of $\hbar\omega$ = 2.3·10$^{-19}$ J and $R$ = 0.32, Eq. (2) gives $E_p$ = 1.1 pJ. This value is by 3-4 times larger under lasing conditions when the emission line is typically shifted by about $kT$ up inside the band from the edge [8]. A spectral hole must appear in the carrier distribution when the electrons and holes recombine simultaneously (see Fig. 4). The maximum peak power is determined by the rate of filling up of this spectral hole with neighbour electrons. It has been demonstrated both experimentally and theoretically [13] that the total time of the filling of the spectral hole in GaAs is about 300 fs. This implies that electrons, which occupy energy levels outside the given 8-meV range, can contribute to the emission power not earlier than after approximately 300 fs. As a result, due to the finite intraband relaxation time the maximum possible peak power of pulses must be not large than around 3.7 W.

It is obvious that experimentally obtained pulse energies and peak powers must be smaller than this maximum value. For comparison, let us note that typical values of the peak power in experiments with similar semiconductor structures were measured to be 0.3 W [14] or 0.6 W [15]. As mentioned above in Section 2, we have experimental values of $E_p$ in a range of 30-40 pJ and peak powers of above 100 W, which are by 30-40 time larger. Hence, the presence of only 2 electrons on each energy level and the distribution of the energy levels $\propto \sqrt{E}$ cannot explain extremely large pulse energy and peak power that we have in experiment.

4. **Optical gain and absorption in the structure**

Consider now qualitatively the behaviour of the optical gain and absorption in a multiple contact structure at different levels of driving current and reverse bias. A typical calculated spectral dependence of the optical gain coefficient in a bulk GaAs/AlGaAs structure is presented in Figure 5 (a) [16]. The electron-hole density is 2·10$^{18}$ cm$^{-3}$, which corresponds to the lowest level of the carrier concentration in our experiments. This value is typical for the laser generation in GaAs/AlGaAs heterostructures with similar parameters [12]. The spectral maximum of the gain lies at the photon energy of 1.424 eV which corresponds exactly to the lasing wavelength detected experimentally [8,9].

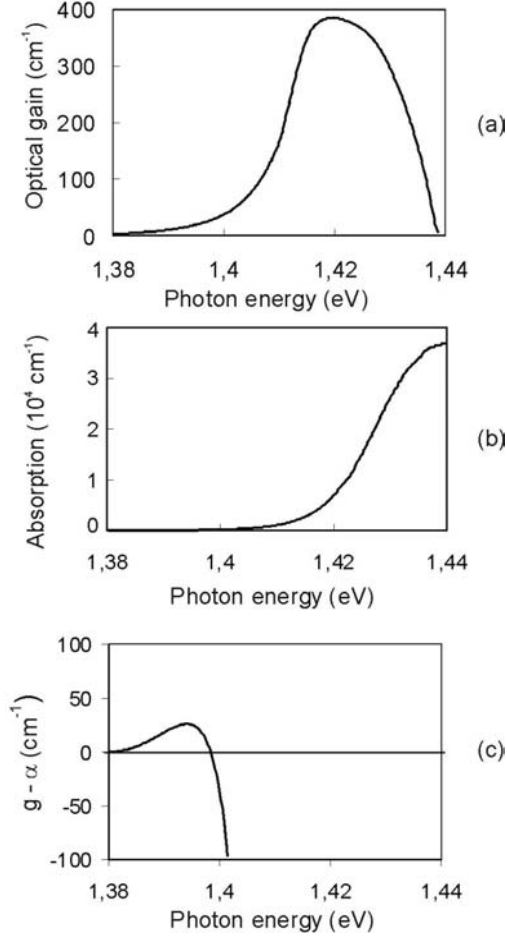

Fig. 5. Optical gain (a), absorption (b) and their difference (c) in a triple-contact GaAs/AlGaAs heterostructure.

The centre of the optical gain shifts towards shorter wavelengths when the carrier density increases and the width of the curve gets broader. Due to the band gap shrinkage at higher concentrations, the long wavelength tail of the gain moves to lower energy region. This tail is well approximated by the dependence $g(\hbar\omega) \propto (\hbar\omega - E_g)^2$.

Just to remind, our structures have tree sections with independent pumping, two of which being amplifiers whereas the last one being an absorber (see [8-11] in detail). The absorption coefficient of the last section $\alpha(\omega)$ depends strongly on frequency and is presented in Figure 5 (b). Comprehensive measurements of $\alpha(\omega)$ at different reverse biases in bulk p-i-n GaAs/AlGaAs heterostructures, which were very similar to our structures, have been carrier out by Knupfer et al [17]. We use here their results. $\alpha(\omega)$ at the long wavelength region is described well by the dependence $\alpha(\omega) \propto \exp\{\alpha_0(\hbar\omega - E_g)\}$, where $\alpha_0$ is a constant. As we noted in Section 2, the absorption edge can be easily shifted by the reverse bias applied to the absorber section (the Frantz-Keldysh effect). It is the strong reverse bias that differs strongly our structures from conventional laser heterostructures. The absorption at the absorber section at photon energies $\hbar\omega$ < 1.42 eV exceeds severely the optical gain in the amplifier sections at reverse biases larger than - 1…- 3 V. The net gain is negative and the lasing is impossible.

The difference $g(\omega) - \alpha(\omega)$ can be readily controlled by simultaneous changing of the driving current and the reverse bias. For instance, Figure 5 (c) illustrates that at some conditions there exists a narrow range at the long wavelengths tail (the excitonic part of the spectrum) where one has transparency or even a small gain. Since the DOS at this region is small, the peak value of the gain is much smaller than the typical lasing gain. The latter has the maximum value shifted by about kT inside the bands. Compare the maximum values in Figs. 5 (a) and (c). The presence of a small gain at the excitonic part of the spectrum plays a decisive role in the condensation of electron-hole pairs and the formation of the coherent BCS-like state in our case.

5. Possible mechanism of the condensation

Figure 6 illustrates a possible condensation mechanism. Photons, which have energies corresponding to the net gain range (see Fig. 5 (c)), can travel freely through the structure and bounce back and forth between the chip facets. The gain is not large enough for lasing.

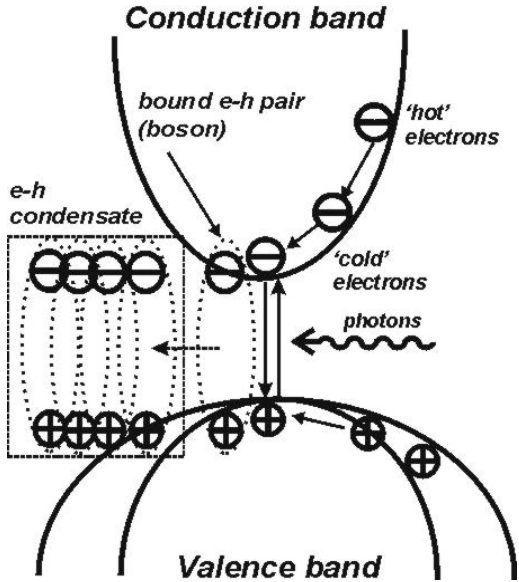

Figure 6. Band structure illustrating the photon mediated drain of the coldest electrons off the fermionic ensemble to the condensate.

The photons create and destroy electron-hole pairs via stimulated emission or absorption. Here, the total wavevector of each pair is approximately equal to zero since the photon wavevector is negligible. Moreover, individual wavevectors of the electron and hole of a pair is also very small because transitions occur between the bottom of the conduction band and the top of the valence band. The electron and hole bounded by a photon becomes a composite boson and it remains coherent with the electromagnetic field and another similar pairs (bosons) for some time. Due to a very fast intraband relaxation rate, electrons and holes from higher energy levels relax towards the bottoms of the band almost simultaneously. Thus, the energy levels of the electron and hole, which has just become the boson, proves to be occupied by another ones. This process is repeated again and again. As a result, the field 'draws off' the coldest *e-h* pairs from the fermionic ensemble to bosonic one where a macroscopically large number is accumulated. Due to a large spontaneous recombination time (> $10^9$ s), the number of the collectively paired electrons and holes can exceed $10^8$.

Let us now consider why the pairs of the condensate are stable for some time until the onset of their cooperative recombination (superradiance). A bound pair can disappear via two different ways. The first way is due to either spontaneous or stimulated emission. The spontaneous lifetime in GaAs is about 2-3 ns, which is very long compared to another processes under consideration. The stimulated emission lifetime is shorter than the previous one and depends on the photon density. However, it is still very long at early stages of the evolution of the cooperative state when the photon density is very small. This way of recombination can be neglected. The second process, which is much more important here, is loss of coherence of the pair due to collisions when the bond between the electron and the hole is being lost. As a result of this process, the composite boson disappears and one has two independent fermions instead.

The paradox consists in the fact that a very large density of *e-h* pairs is required for stability of the condensate and suppression of the ultrafast relaxation rate due to collisions. Let us demonstrate this by comparing the distribution of electrons in the conduction band at two densities corresponding to lasing and the cooperative emission. A typical density of *e-h* pairs required for lasing is in a range of $(1-2) \cdot 10^{18}$ cm$^{-3}$, whereas it lies between $(3-6) \cdot 10^{18}$ cm$^{-3}$ in the superradiance regime [8,9]. Figure 7 shows the Fermi-Dirac distribution of electrons in GaAs at two densities of electrons. The upper plot corresponds to the electron distribution at the *e-h* density of $1.5 \cdot 10^{18}$ cm$^{-3}$, the renormalised band gap $E_g = 1.4095$ eV, and the Fermi energy $\mu = 60.7$ meV (lasing). The lower plot corresponds to the cooperative emission at the density of $6 \cdot 10^{18}$ cm$^{-3}$, $E_g = 1.3946$ eV, and $\mu = 172.1$ meV. The arrows show the position of the band gap, the quantum energy $h\nu$, and the Fermi energy in each case. A value of $(1-f_e)$, which is equal to a probability of finding empty the level with the energy *E*, is also shown in the lower plot for illustration. The energy of the optical phonon $\hbar\Omega$ (= 36 meV) is plotted as well.

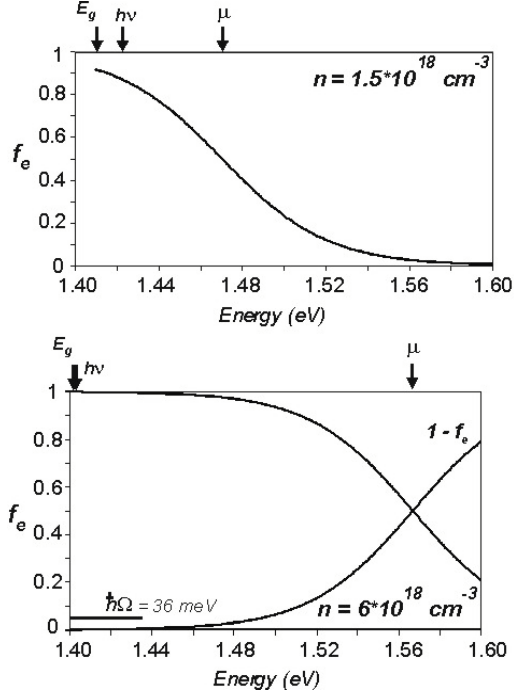

Figure 7. The Fermi-Dirac distribution of electrons in GaAs at room temperature for two values of density.

The lasing wavelength locates about 20-25 meV inside the band. The probability of occupation of the working levels for an electron is around 0.85-0.87. There are a lot of free levels both above and below the working levels. *e-h* pairs, bound by the electromagnetic field and coherent with it (bosons), are dephased and destroyed by colliding with neighbour electrons and holes. Therefore, the electron and hole of a bound pair becomes independent fermions. For all this, the electron of the bound pair can easily find a free energy level where to go. That is why the speed of the pair destruction is very fast and the accumulation of bosons does not happen. The number of *e-h* pairs that are coherent with the electromagnetic field is very small (about $10^{-3}$ - $10^{-4}$) [11,18].

A totally different picture is observed in the large density case (see the lower plot in Fig. 7). The peak wavelength of the cooperative emission locates a few meV only above the band gap edge. The Fermi energy is far inside the conduction band at such a high density (a highly degenerate semiconductor). As a result, practically all the energy levels within a 30-50 meV wide zone above the band edge are occupied by electrons. This implies that a bound *e-h* pair is not destroyed and dephased by collisions with its neighbours because there is no place for the electron of the pair where to go (the Pauli exclusion principle). The only possibility is to occupy a free energy level far inside the conduction band. But the probability of such a transition would be very small since the energy and wavevector mismatches must be compensated somehow. Even the energy of the optical phonon is 36 meV only. On other words, bound *e-h* pairs are squeezed between the band gap from below and the occupied subband from above as shown in Fig. 8. The occupied subband expands with an increase of the *e-h* density and *vice versa* it shrinks while the number of free electrons decreases when they get gradually into the bosonic ensemble (*e-h* BCS-like state). When the majority of electrons transforms into composite bosons and the number of unbound electrons decreases dramatically, the occupied subband shrinks so much that the relaxation and destruction of bosons into components will be possible again. The process of the formation of the bosonic subsystem stops self-consistently.

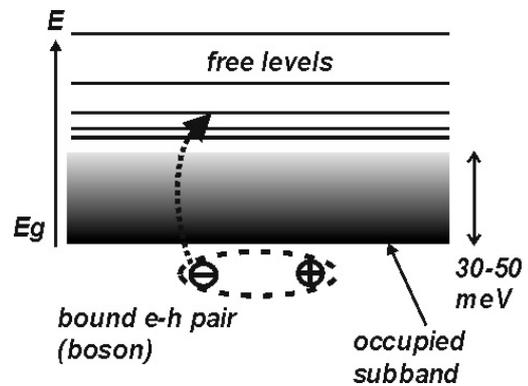

Figure 8. The subband occupied by electrons which prevents the distruction of the condensate and loss of its coherency.

A possible analogy is as follows. The occupied subband in our case plays a similar role as the energy gap in a superconductor that prevents the destruction of Cooper pairs at temperatures below the critical one. The obvious difference is that all the levels in the superconducting gap are banned for electrons, whereas all the energy levels in the subband are occupied by electrons. In both cases transitions of electrons there are impossible.

In contrast to the lasing case, the number of bound *e-h* pairs that are coherent with the electromagnetic field is very large and can approaches to the total number of electrons and holes injected into the structure [11]. The cooperative BCS-like state finally recombines and forms a femtosecond giant pulse. This explains the record level of peak powers of superradiant pulses observed experimentally.

The proposed mechanism allows us to explain all the peculiarities of the cooperative emission in GaAs, including the anomalously long coherence time, the extremely small Fermi energy (6-8 meV) of electrons of the coherent BCS-like state [8-10]. Moreover, the transformation of the DOS electrons of this state explains very large power levels of bandwidth-limited femtosecond superradiant pulses.

6. **Conclusions**

In conclusion, the mechanism explaining the observation of the *e-h* pairs condensation and the formation of a coherent *e-h* BCS-like state in bulk GaAs at room temperature has been presented. We have shown that experimentally obtained peak power and energy of superradiant pulses cannot be explained by a simple square-root dependence of DOS in the bands and Fermi-Dirac energy distribution of electrons in GaAs.

We have demonstrated that the simultaneous fulfilment of two conditions is crucial. The first condition is the presence of a resonant electromagnetic field which can assist the pairing of electrons and holes located at the bottom of the conduction band and top of the valence band. The second is a large *e-h* density, much larger than typical densities required for lasing. A technique for the achievement of such carrier concentrations has been discussed elsewhere [10]. A high extent of degeneracy of the semiconductor at very large densities results in bringing about of a relatively broad (30-50 meV) occupied electronic subband, which prevents the destruction of the *e-h* condensate and loss its coherency with the photon field due to collisions. This mechanism has been demonstrated in multiple-section (amplifier/absorber) *p-i-n* GaAs/AlGaAs heterostructure. However, in our view, it has a rather general behaviour and may be used for the achievement of the condensed *e-h* state in another semiconductors.

Finally, the author would like to thank H. Kan, H. Ohta, and T. Hiruma of Hamamatsu Photonics K.K. for the financial support.